\documentclass[prl,aps, showkeys, showpacs]{revtex4}
\def\b{\begin{equation}}
\def\e{\end{equation}}

\begin{document}

\title{Quantum Information Paradox: Real or Fictitious?}

\author{Abhas Mitra}

\email {amitra@barc.gov.in}
\affiliation { Theoretical Astrophysics Section, Bhabha Atomic Research Centre, Mumbai, India}
%\affiliation %Theoretical Astrophysics Section, Bhabha Atomic
%Research Center, Mumbai-400085, India}

 %\altaffiliation[Presently at ]{Theoretical Astrophysics Section, Bhabha Atomic Research Centre, Mumbai, India}%Lines break %automatically or can be forced with \\

\date{\today}% It is always \today, today,


%\affiliation %Theoretical Astrophysics Section, Bhabha Atomic
%Research Center, Mumbai-400085, India}

 %\altaffiliation[Also at ]{Physics Department, XYZ %University.}%Lines break %automatically or can be forced with \\



\date{\today}% It is always \today, today,
\begin{abstract}
One of the outstanding puzzles of theoretical physics is whether quantum
information indeed gets lost in the case of Black Hole (BH) evaporation or accretion. Let us
recall that Quantum Mechanics (QM) demands an upper limit on the acceleration of
a test particle. On the other hand, it is pointed out here that, if a
Schwarzschild BH would exist, the acceleration of the test particle would {\em blow
up at the event horizon in violation of QM}. Thus the concept of an exact BH is
in contradiction of QM and quantum gravity (QG). It is also reminded that the mass of a BH actually
appears as an INTEGRATION CONSTANT of Einstein equations. And it has been shown that
the value of this integration constant is actually zero! Thus even classically,
there cannot be finite mass BHs though zero mass BH is allowed. It has been
further shown that during continued gravitational collapse, radiation emanating
from the contracting object gets trapped within it by the runaway gravitational
field. As a consequence, the contracting body attains a quasi-static state where
outward trapped radiation pressure gets balanced by inward gravitational pull and  the
ideal classical BH state is never formed in a finite proper time. In other words, continued gravitational collapse results in an ``Eternally Collapsing Object'' which is a ball of hot plasma and which is asymptotically approaching  the true BH state with $M=0$ after radiating away its entire mass energy. And if we
include QM, this contraction must halt at a radius suggested by highest QM
acceleration. In any case no EH is ever formed and in reality, there is no
quantum information paradox.

%\pacs{04., 04.20.Jb, 04.20.Cv, 03.50.-z}
%\verb+\pacs{#1}+ command.
\end{abstract}
%\pacs{04.02.-q, 04.02.Cv, 04.04.-b, 04.30.-w}% PACS, the Physics and Astronomy
%\keywords {Classical general relativity,  Classical Field Theory}%Use showkeys class option if keyword


\maketitle                        %display desir

% 
% May 25, minor revision
%\documentstyle[pramana,epsf,floats]{ias}
%\documentstyle[pramana,epsf,floats,amsmath, amsthm, amsfonts, amssymb, bbm,mathrsfs,graphicx]{ias}
%\def\b{\begin{equation}}
%\def\end{equation}{\end{equation}nd{equation}}
%\def\l{\left}
%\def\r{\right}
%\def\l{\left}
%\def\r{\right}
%\documentstyle[12pt,epsfig]{article}
%\draft
%\usepackage{epsfig}
%\usepackage{subeqn}
%\usepackage{amsmath}
%\renewcommand{\baselinestretch}{1.3}
%\textwidth 17cm
%\textheight 23. cm
%\oddsidemargin -.5 cm
%\end{equation}vensidemargin 1 cm
%\topmargin -.5 cm

\vskip 5mm
\noindent

\section{Introduction: Upper Limit on Acceleration in Quantum Gravity}
Quantum Gravity imposes a fundamental unit of {\em proper} length called {\em Planck Length}:
\begin{equation}
l_p = \sqrt{{\hbar G\over c^3}}
\end{equation}
and a fundamental unit of  time called Planck Time $\tau_p = l_p/c$. Accordingly any QG should impose a Maximal Acceleration:
\begin{equation}
a_p = {Maximum ~~Speed \over  Minimum ~~Proper ~~Time} \sim {c^2 \over
l_P} = {c^{7/2} \over {\sqrt{\hbar G}}}
\end{equation}

Detail considerations yields\cite{1,2}
\begin{equation}
a_p =  {c^2 \over
2 l_P} = {c^{7/2} \over \ 2 {\sqrt{\hbar G}}}
\end{equation}
Thus $QM$ {\em does not allow occurrence of infinite proper acceleration}. Some preliminary formulation of QG indeed mentions of Maximal Acceleration\cite{3}.

\subsection{ Proper Acceleration Due to Gravity}
The {\em Invariant Proper} Acceleration on a test particle around lying on the surface of a spherical body of mass $M$ and Radius $R_0$\cite{4}
\begin{equation}
a = \sqrt{-a^i a_i} = {GM\over R_0^2\sqrt{1 - 2GM/R_0c^2}}
\end{equation}
And if the body is instead a Schwarzschild black hole with $R_0 = {2 G M/ c^2}$, one would have $a = \infty$\cite{4}!! Note that since $a$ is an {\em invariant/scalar}, its value is independent of coordinates and even a free falling test particle would experience
{\em infinite acceleration}. Clearly, such a situation would be in violation of the inherent concept of a natural QG upper limit of acceleration $a_p$. Thus no meaningfully formulated QG should admit existence of BHs. Hence in properly formulated QG theories, there should be neither any horizon, nor any BH, nor any Hawking Radiation. Thus as per QG, either there must be some physical process to stop continued gravitational collapse at a certain $R_0=R_p$ where $a \le a_p$ or else, in the case of Schwarzschild BHs, the mass of the BH should be zero: $M =0$. This is so because of two reasons:

$\bullet$ The proper time of formation of a zero mass BH is infinite: $\tau_{formation} \propto M^{-1/2} =\infty$.

$\bullet$ Even if mathematically, one would {\em assume} the existence of such a zero mass BH, the proper time of infall of a test particle
for reaching the Event Horizon (EH) $\tau_{infall} \propto M^{-1/2} =\infty$.

In this way, the conflict of avoiding an infinite acceleration would be avoided.
\subsection{Hint From String Theories}
There are some solutions in String Theories which do not admit any horizon or singularity. For instance 

$\bullet$ ``If one applies the transformation (2.13) to time transformation in Schwarzschild, one obtains a solution in which {\em the horizon
becomes a singularity}.''\cite{5}

Note, horizon itself becomes the singularity only when mass of the BH : $M =0$ because radius of the BH, $R_g = 2M$ ($G=c=1$).

$\bullet$ ``D1-D5 solutions have neither any horizon nor
any singularity.''\cite{5}

$\bullet$ ``Straight Strings have no  Event Horizon''\cite{5}.

Some solutions of the string theories specifically suggest that point particle are massless:

$\bullet$ ``The consistency of our picture requires that there is one and only one supermultiplet which becomes massless at the singularity''\cite{6}

``At first sight it may seem surprising that classical black holes can be massless. However this phenomenon has an appealing explanation from 10-D
perspective. The IIA(IIB) theory has extremal black twobrane (threebrane) solutions whose mass is proportaional to their area. After Calabi- Yau compactification these may wrap around minimal two (theree) surface and appear as a 4-D black hole. As the area of the surfaces around which they wrap goes to zero, the corresponding BH becomes {\em massless}....''\cite{6}.

When BH is massless, its horizon area $A =0$, and this a unique state with entropy
\begin{equation}
S = \ln 1 =0
\end{equation}
Physically a massless  BH with zero horizon area confines no information. There is no question of evaporation or information loss in such a case. Nor does a test
particle ever reach the horizon of a massless BH $\tau_{infall} \propto M^{-1/2} = \infty$.

In fact, there are specific indications that QG does not actually allow any Hawking Radiation (even if the existence of a BH would be assumed:

``A New argument is presented confirming the point of view that a Schwarzschild BH formed during a gravitational collapse process does {\bf NOT}
radiate''\cite{7}.

$\bullet$ ` Hawking effect may be only a mathematical artifact because it demands singularity of the wave function at event horizon in violation of QM..''\cite{8}.

\section{ Does General Relativity Allow Finite Mass BHs?}
The existence of the vacuum Schwarzschild solution apparently suggests existence of BHs of arbitrary mass. But what is forgotten here is that
(i) The ``mass'' of the BH appears through an {\bf integration constant}: $\alpha = 2 GM/ c^2$ or $M = {c^2 \alpha/ 2G}$.

(ii) Although symbolically ``integration constants'' may {\em look} to assume arbitrary value, in reality, some of them may have a {\em definite} or precise value.

\subsection{ Fixing the value of this Integration Constant}

One may consider the exterior Schwarzschild BH solution in two different coordinates: 

$\bullet$  The original Schwarzschild Solution : 
\begin{equation}
ds^2 = (1-\alpha/R) dT^2 - (1-\alpha/R)^{-1} dR^2 - R^2 (d\theta^2 + \sin^2 \theta d\phi^2)
\end{equation}
$\bullet$  The Eddigton -Finkelstein Solution\cite{9,10,11,12}: 
\begin{equation}
ds^2 = (1-\alpha/R) dT_*^2  \pm (2\alpha/R) dT_* dR - (1+\alpha/R) dR^2 - R^2 (d\theta^2 + \sin^2 \theta d\phi^2)
\end{equation}
 where 
 \begin{equation}
 dT_* = dT + {\alpha \over R -\alpha} dR
\end{equation}
and apply the condition that proper 4-volume remains invariant for any coordinate transformation, i.e.,
\begin{equation}
 \sqrt{-g_*}~ dR ~dT_*~ d\phi~ d\theta = \sqrt{-g} ~dR~dT ~d\theta ~d\phi
 \end{equation}
 where $g = \det |g_{ik}|$ and happens to be same in both the cases: $g = g_*=- R^4 \sin^2 \theta$. Then the foregoing Eq. yoields
 $dT_* = dT$ which, by virtue of Eq.(98 demands
 \begin{equation}
 \alpha = {2GM\over c^2} =0
 \end{equation}
 or $M=0$. This means that though the mass of a Schwarzschild BH which appears to be arbitrary, {\em actually, it is unique}: $M \equiv 0$! It is important to note that if instead of a ``point particle'' we would be considering a {\em finite}  static spherical body of gravitational mass
$M$, the above derivation would be invalid even though by Birchoff's theorem, the exterior spacetime would still be described by metric (2).  This happens because the coordinate transformation(8)
is obtained by integrating the {\em vacuum} null geodesic all the way from $R=0$. And this is allowed only when the spacetime is indeed due to a point mass and not due to a finite body. For a finite body, the {\em interior metric would be different} from what is indicated by Eqs.(6) and (7).
 As explained above, $M=0$ BHs are never formed, even if they would be assummed to be there, there is no Hawking Radiation!
 
 More importantly, this means that the observed BH Candidates having finite massess cannot be exact BHs though they may resemble 
 Dark/Black like theoretical BHs. This would be so if the Gravitational Redshft of the real BH Candidates, $z \gg 1$.
 WHY? This is so because observed luminosity of an object falls off as
 \begin{equation}
 L_{observed} \sim (1+z)^{-2} \to 0~as~z\gg 1
 \end{equation}
 where one defines
 \begin{equation}
 z = (1- 2GM/Rc^2)^{-1/2} -1
 \end{equation}
 Recall here that for ideal BHs, $z=\infty$ and consequently $L_{observed} = 0$.
 \section{Consistency With Gravitational Collapse}
 The metric of a spherically evolving fluid is given by
 \begin{equation}
ds^2 = g_{00} dt^2 + g_{rr} dr^2 - R^2 (d\theta^2 + \sin^2\theta d\phi^2)
\end{equation}
where $R=R(r, t)$ is the Invariant Circumference coordinate (also
called as areal coordinate) and happens to be
 a scalar. Further,  for radial motion with $d\theta =d\phi =0$,
the metric becomes
\begin{equation}
ds^2 = g_{00} dt^2 (1- x^2)
\end{equation}
where
\begin{equation}
 x = {\sqrt {-g_{rr}} ~dr\over \sqrt{g_{00}}~ dt}
 \end{equation}
 We may recast Eq.(14) as
\begin{equation}
(1-x^2) = {1\over g_{00}} {ds^2\over dt^2}
\end{equation}
 The comoving observer at $r=r$ is
free to do measurements of not only the fluid element at $r=r$ but also of other objects:
  If the
comoving observer is compared with a static floating boat in a
flowing river, the boat can monitor the motion of other boats or the
pebbles fixed on the river bed. Here the fixed markers on the river
bed are like the background $R=  fixed$ markers against which the
river flows. Alternatively, the comoving observer may be seen as the
driver of a car (local fluid) while the fixed milestones on the
roadside constitute the $R=fixed=scalar$ grid naturally definable
for any spherically symmetric problem. If we intend to find the
parameter $x$ for such a $R=constant$ milestone,  we will have, $dR
=0$, or
\begin{equation}
 d R(r,t) = 0= {\dot R} dt + R^\prime dr
 \end{equation}
where an overdot denotes a partial derivative w.r.t. $t$ and a prime denotes
a partial derivative w.r.t. $r$.
Therefore for the $R=constant$ marker, we find that
\begin{equation}
{dr\over dt} = - {{\dot R}\over R^\prime}
\end{equation}\footnote{Since here $|dr/dt|=|{\dot R}/R^\prime|$; partial derivatives effectively behave like total ones. Hence, the allegation by Kundt (arXiv:0905.1028)  that I confused partial derivatives with total ones is incorrect. He overlooked that I was considering {\em noncomoving}  $R=const$ observers.}
and the corresponding $x=x_c$ is
\begin{equation}
x= x_{c} = {\sqrt {-g_{rr}} ~dr\over \sqrt{g_{00}}~ dt} = -{\sqrt {-g_{rr}}
~{\dot R}\over \sqrt{g_{00}}~ R^\prime}
 \end{equation}
Using Eq.(16), we also have
\begin{equation}
(1-x_c^2) = {1\over g_{00}} {ds^2\over dt^2}
\end{equation}
Now let us define\cite{13,14,15} two auxiliary parameters
\begin{equation}
\Gamma = {R^\prime\over \sqrt {-g_{rr}}}; \qquad U = {{\dot R}\over \sqrt{g_{00}}}
\end{equation}
so that the combined Eqs. (18), (19) and (21) yield
\begin{equation}
x_c = -{U\over \Gamma}; \qquad U= -x_c \Gamma
\end{equation}
As is well known, the gravitational mass of the collapsing (or
expanding) fluid is defined through the equation\cite{4,13,14,15}
\begin{equation}
\Gamma^2 = 1 + U^2 - {2M(r,t)\over R}
\end{equation}
Then the two foregoing equations can be combined and transposed to obtain
\begin{equation}
\Gamma^2 (1- x_c^2) = 1- {2M(r,t)\over R}
\end{equation}
By using Eqs.(20) and (22) in the foregoing Eq., we obtain
\begin{equation}
{{R^\prime}^2\over {-g_{rr} g_{00}}} {ds^2\over dt^2} = 1 - {2M(r,t)\over R}
\end{equation}
Recall that the determinant of the metric tensor is always
negative\cite{4,13,14,15}:
\begin{equation}
g = R^4 \sin^2 \theta ~g_{00} ~g_{rr} \le 0
\end{equation}
so that we must always have
\begin{equation}
-g_{rr}~ g_{00} \ge 0
\end{equation}
As mentioned before, all worldlines of  material particles or observers must be timelike
because the comoving metric (or for that matter even the appropriate Schwarzschild metric) has no supposed
``coordinate singularity'' unlike the vacuum Schwarzschild case. 
For the signature chosen here,  this means, $ds^2 > 0$ for all observers.
 Then by noting Eq.(27), it follows that the
LHS of  Eq.(25) is {\em always positive}. So must then be the RHS of
the same Eq. and which implies that
\begin{equation}
{2M(r,t)\over R} <1 
\end{equation}
This shows in the  utmost general fashion that trapped surfaces
are not formed in spherical collapse or expansion\cite{4,13,14,15,16}. 
 Thus the crucial assumption behind Singularity Theorems of Hawking, Penrose, Geroch, i.e., ``Formation of a Trapped Surface''
 {\em is actually not realized} in General Relativity. Therefore, if a fluid would indeed collapse to a singularity at $R\to 0$, one must have $M \to 0$ in order that the inequality is honoured. And this is consistent with the result that for BHs, $\alpha = 2GM/Rc^2 =0$ and BHs have unique mass $M\equiv 0$!
 \section{ Eternally Collapsing Objects}
 Two natural questions here would be: 
   1. What about the upper limit of White Dwarf and Neutron Star Masses?
  
  Answer: These  limits are intended for objects supported by {\bf cold} quantum degeneracy pressure. On the other hand there is NO UPPER LIMIT
  for HOT objects supported by RADIATION PRESSURE\cite{17,18,19,20}.
  
  2. If there cannot be any finite mass BH what are the true nature of the observed BH Candidates?
  
  Answer: It has been showed that during continued gravitational collapse, the radiation (photons/neutrinos) emanating from the collapsing object
  would get almost trapped by the runaway gravitational field. Finally a state would be reached when\cite{17,18,19,20,21}:
  \begin{equation}
  Inward ~pull ~of~gravity = Outward ~trapped~radiation ~pressure
  \end{equation}
   and consequently, the collapsing object becomes a hot ball of quasistatic plasma where extreme radiation pressure balances the gravity. However, it is still a quasi-static stage and is contracting at infinitesimally slow rate to attain the exact BH state with $M=0$. It has also been shown
  that this contraction process continues indefinitely and hence the BH Candidates are Eternally Collapsing Objects (ECOs) rather than true BHs!
  \section{Observational Prediction and Verification}
  As an astrophysical plasma contracts, the frozen in magnetic field increases as $B \propto R^{-2}$ and this is the basic reason that young neutron stars have $B \sim 10^{12-13}$ G. Further in the extreme relativistic stages with $z\gg 1$, {\em local} value of $B$ may additionally increase as $B\propto (1+z)$. On this basis, the present author had predicted that the observed BH Candidates, i.e., ECOs will have strong {\em intrinsic} magnetic field\cite{4,13,20}. In contrast {\em ture chargeless BHs have no intrinsic magnetic field}.
  Consequently, some of the properties
  as well as accretion geometries around  ECOs could resemble, in some way to those of magnetized Neutron  Stars rather than (true) BHs.
 And this prediction has been verified  both for stellar mass (galactic) and extragalactic BH Candidates (i.e., ECOs)\cite{22,23,24,25,26}. In particular, for the extragalactic cases, microlensing observations spanning 20 years have given almost direct proof that, the some of the Quasars, in particular $Q0957+561$,  harbour Magnetized ECOS (MECOS) rather than
 true BHs\cite{25,26}.

 \section{Conclusion}
  Though BH solution is a correct an exact solution, the integration constant defining the mass of a Schwarzschild BH has the unique value
 $M=0$. This means that the exact BH solution represents only an ideal state which can be attained asymptotically during continued gravitational
 collapse. At any finite proper time, the collapsing object is never, an ideal, exact BH. On the other hand, in the context of classical General Relativity, the actual BH Candidates are likely to be (magnetized) ECOs, a ball of relativistic plasma where inward pull of gravity is balanced by
 outward radiation pressure. Since all meaningful Quantum Gravity theories must reduce to GR in the low energy regime, the apparent BH like solutions
 too must be pointing only ideal limiting/asymptotic  states which can never exactly occur in Nature. Since there is no exact BH, no Event Horizon, there is no Hawking Radiation and Quantum Information Paradox. As mentioned in the beginning, the existence of an upper limit of a finite proper acceleration
 $a_p$ directly contradicts the notion of exact BHs.
 
 In an important paper, it was shown that if BHs would be assumed to be present, Quantum Information Paradox would be even more severe than
 usually considered\cite{27}. However, rather than noting that, the most physical solution to this severe problem would lie in the realization that there cannot be finite mass BHs
or Hawking Radiation, the authors  suggested  far fetched possibilites. But it is hoped that it would be  realized that Quantum Information paradox is actually non-existent.
%\section*{References:}


\begin{thebibliography}{99} 

%\begin{equation}ibitem{long}
%\begin{equation}egin{enumerate}
\bibitem{1} Arun Pati, Euro Phys Lett., {\bf 18} (1992), 285.
 \bibitem{2} Arun Pati, IL Nuovo Cimento B, {\bf 107} (1992), 895.
\bibitem{3}  E.R. Caianiello et al., Int. J. Mod. Phys. D, {\bf 3(2)} (1994), 485.
\bibitem{4} A. Mitra,  Found. Phys. Lett., {\bf 15(5)} (2002), 439.
 
\bibitem{5} G. Horowitz,  (1992) (hep-th/9210119)

\bibitem{6} A. Strominger, Nucl. Phys. B., {\bf 451} (1995), 109,  (hep-th/954090).
\bibitem{7}  V.A. Belinski, Phys. Lett. A {\bf 354} (2006), 249, gr-qc/060713.

 \bibitem{8} M. Rabinowitz, (2004);  astro-ph/0412101.

\bibitem{9} A. Mitra, Adv. Sp. Res.,  {\bf 38(12)} (2006), 2917, (astro-ph/0510162).
\bibitem{10} A. Mitra, J. Math. Phys. {\bf 50 (4)},  (2009), arXiv:0904.4754
\bibitem{11} A. Mitra (2005), physics/0504076.
\bibitem{12} A. Mitra, Proc. 11th Marcel Gross Conf. on General Relativity (2006, Berlin, World Scientific)
 
\bibitem{13} A. Mitra, Found. Phys. Lett., {\bf 13(6)} (2000), 543.
\bibitem{14} A. Mitra, (2004), astro-ph/0408323.
 \bibitem{15} A. Mitra, (2005)  gr-qc/0512006.
\bibitem{16} D. Leiter \& S.R. Robertson, Found. Phys. Lett., {\bf 16} (2003), 143.
 
 \bibitem{17} A. Mitra, Mon. Not. R. Astron. Soc. Lett., {\bf 367} (2006), L66; gr-qc/061025.
  
  
  \bibitem{18}  A. Mitra,  Mon. Not. R. Astron. Soc., {\bf 369} (2006), 492; gr-qc/0603055.
  
  \bibitem{19} A. Mitra, Phys. Rev. D {\bf 74(2)} (2006), 02010; gr-qc/0606066.
  
  \bibitem{20}  A. Mitra, New Astronomy, {\bf 12(2)} (2006), 146; astro-ph/0608178.
  \bibitem{21} A. Mitra \& N.K. Glendenning,  	eScholarship Repository, Lawrence Berkeley National Laboratory, University of California (2006),  	 see, http://repositories.cdlib.org/lbnl/LBNL-59320.
 \bibitem{22} S. Robertson and D. Leiter,  Astrophys. J., {\bf 565} (2002), 447.
  
  \bibitem{23}  S. Robertson and D. Leiter, Astrophys. J., {\bf 569} (2003), L203.
  
 \bibitem{24} S. Robertson and D. Leiter,  Mon. Not. R. Astron. Soc., {\bf 350} (2004), 1391.
 
 \bibitem{25} R. Schild, D. Leiter \& S. Robertson, 	 Astronomical J.,  {\bf 132(1)} (2006),  420.


\bibitem{26}  R. Schild, D. Leiter \& S. Robertson ,  Astronomical J., {\bf 135(3)} (2008), 947.
 \bibitem{27} Arun Pati and S.L. Baurstein, Phys. Rev. Lett.,  {\bf 98(8)} (2007),  080502. 
 
%\end{equation}nd{enumerate}	

\end{thebibliography}
\end{document}